\documentclass[sigconf]{acmart}
\usepackage{multirow}

\AtBeginDocument{%
  \providecommand\BibTeX{{%
    \normalfont B\kern-0.5em{\scshape i\kern-0.25em b}\kern-0.8em\TeX}}}

\setcopyright{acmcopyright}
\copyrightyear{2022}
\acmYear{2022}
\setcopyright{rightsretained}
\acmConference[WWW '22]{Proceedings of the ACM Web Conference
2022}{April 25--29, 2022}{Virtual Event, Lyon, France}
\acmBooktitle{Proceedings of the ACM Web Conference 2022 (WWW '22),
April 25--29, 2022, Virtual Event, Lyon,
France}\acmDOI{10.1145/3485447.3512138}
\acmISBN{978-1-4503-9096-5/22/04}

\begin{document}
\sloppy
\title{How Do Mothers and Fathers Talk About Parenting to Different Audiences?}
\subtitle{Audience Effects and Stereotypes: An Analysis of r/Daddit, r/Mommit, and r/Parenting Using Topic Modelling}
%%
%% The "author" command and its associated commands are used to define the authors and their affiliations.
\author{Melody Sepahpour-Fard}
\orcid{0000-0002-8472-9514}
\affiliation{%
  \department{SFI Centre for Research Training in Foundations of Data Science and Department of Mathematics and Statistics}
  \institution{University of Limerick}
  %\streetaddress{Sarsfield Ave}
  \city{Castletroy}
  \state{Limerick}
  \country{Ireland}
  \postcode{V94 T9PX}
}
\email{Melody.SepahpourFard@ul.ie}

\author{Michael Quayle}
\orcid{0000-0002-7497-0566}
\affiliation{%
  \department{Centre for Social Issues Research and Department of Psychology}
  %\streetaddress{Plassey Park Rd}
  \institution{University of Limerick}
  \city{Castletroy}
  \state{Limerick}
  \country{Ireland}
  \postcode{V94 T9PX}
}
\affiliation{%
  \department{Department of Psychology, School of Applied Human Sciences}  
  \institution{University of KwaZulu-Natal}
  %\streetaddress{Mazisi Kunene Rd}
  \city{Durban}
  \state{KwaZulu-Natal}
  \country{South Africa}
  \postcode{4041}
}
\email{Mike.Quayle@ul.ie}


\begin{abstract}
While major strides have been made towards gender equality in public life, serious inequality remains in the domestic sphere, especially around parenting. The present study analyses discussions about parenting on Reddit (i.e., a content aggregation website) to explore audience effects and gender stereotypes. It suggests a novel method to study topical variation in individuals' language when interacting with different audiences. Comments posted in 2020 were collected from three parenting subreddits (i.e., topical communities), described as being for fathers (r/Daddit), mothers (r/Mommit), and all parents (r/Parenting). Users posting on r/Parenting and r/Daddit or on r/Parenting and r/Mommit were assumed to identify as fathers or mothers, respectively, allowing gender comparison. Users' comments on r/Parenting (to a mixed-gender audience) were compared with their comments to single-gender audiences on r/Daddit or r/Mommit using Latent Dirichlet Allocation (LDA) topic modelling.

Results showed that the most discussed topic among parents is about education and family advice, a topic mainly discussed in the mixed-gender subreddit and more by fathers than mothers. The topic model also indicated that, when it comes to the basic needs of children (sleep, food, and medical care), mothers seemed to be more concerned regardless of the audience. In contrast, topics such as birth and pregnancy announcements and physical appearance were more discussed by fathers in the father-centric subreddit. Overall, findings seem to show that mothers are generally more concerned about the practical sides of parenting while fathers' expressed concerns are more contextual: with other fathers, there seems to be a desire to show their fatherhood and be recognized for it while they discuss education with mothers. These results demonstrate that concerns expressed by parents on Reddit are context-sensitive but also consistent with gender stereotypes, potentially reflecting a persistent gendered and unequal division of labour in parenting.

\end{abstract}




\maketitle

\section{Introduction}
The present study explores how mothers and fathers express concerns related to parenting on Reddit. Using topic modelling, more than a hundred thousand comments published on Reddit in 2020 were analysed to describe parenting talk. Parents' comments on three subreddits related to parenting---two single-gender (r/Daddit and r/Mommit) and one mixed-gender (r/Parenting)---were analysed using topic modelling to see whether speakers are sensitive to the audience's gender identity and if the content of parents' talks reflects gender stereotypes. Parenting is an important domain of study because women frequently face a double burden of working full-time outside the house but also carrying the primary responsibility for caring and home-making \cite{brines_second_1990}. Consequently, women often lower their professional ambitions to satisfy the needs at home and, therefore, fragilise their financial independence \cite{raley_when_2012}. Equality in the public sphere first requires equality at home.
Why are women still preferentially perceived to be primary caregivers \cite{wall_how_2007}? Generalisations about the shared attributes of social groups \cite{judd_definition_1993}, or stereotypes, constitute an essential factor in assigning such duties to women \cite{ellemers2018gender}. From simple beliefs, stereotypes become concrete in shaping individuals' behaviour once they are widely accepted and become injunctive norms for a social group to act in a specific way \cite{ellemers2018gender}.

The parenting domain has especially profoundly rooted stereotypes about how mothers and fathers should behave. For this reason, the present study will focus on discourses related to parenting behaviours using congruency with gender stereotypes as a framework of understanding. In the literature review, we will first define stereotypes related to gender and show how they can create expectations for mothers and fathers regarding parenting behaviours. Secondly, we will integrate audience as a factor that can make a social category salient and facilitate identity-relevant behaviour. For this second part, we will define social categorisation, how it has been theorised, and its relation with social identity performance \cite{klein_social_2007} in front of an audience. Finally, we will describe the online context of Reddit, the selected subreddits, and related previous research.

\section{Literature Review}
\subsection{Gender Stereotypes and Parenting}
Gender stereotypes assign specific attributes, characteristics, and responsibilities, and these become standards to which individuals should conform to be positively judged by society \cite{villicana_gender_2017}. Generally, women are assigned "communal" characteristics such as selflessness and care, and men "agentic" characteristics such as mastery and assertiveness \cite{eagly_gender_1984}. These gender stereotypes are associated with the domains of family for women and work for men, and thereby culturally facilitate home-making for women and outside work for men \cite{ellemers2018gender}. Put differently, these gendered associations can undermine women's professional ambitions in favour of their domestic identity. For instance, women are more likely than men to pass up promotions, take less demanding jobs, switch to part-time jobs, and drop out of the labour force once they become parents \cite{raley_when_2012}. The reverse is true for fathers, who are still considered secondary parents \cite{wall_how_2007}, even if they aspire to be more involved in caring \cite{bulanda_paternal_2004}.

\subsection{Social Categorisation and Audience Related to Stereotypical Behaviours}
\subsubsection{Social Categorisation and its Theorisation}
Categorisation – the assignment of people to discrete categories based on fuzzy criteria – is a fundamental psychological process \cite{tajfel_social_1974, tajfel_integrative_1979}, resulting in perceptions of groups characterised by sets of defining characteristics, or stereotypes \cite{ellemers2018gender}. Social categorisation is, therefore, a precondition to the formation of stereotypes. As described in the Social Identity Theory \cite{tajfel_social_1974, tajfel_integrative_1979}, when people identify with a social group, they think, feel, and act as group members \cite{tajfel_integrative_1979}, and appropriate behaviour is defined by the group's self-stereotypes and norms \cite{tajfel_integrative_1979}. The Self-Categorization Theory  \cite{turner_rediscovering_1987} adds a dynamic aspect, noting that people can shift across social identities as they move across social contexts. When individuals define themselves with a social category in a specific context, they tend to conform to the norms characterising their social identity in that context \cite{turner_social_1991}.

"Mother", "father", and "parent" are social categories individuals can identify with. They can become salient in a given context and prompt individuals towards more identity-relevant (i.e., stereotypical) behaviours. Thus, the audience is an important trigger of the salience of social categories.

\subsubsection{Audience and Self-Categorisation}
The social identity performance framework \cite{klein_social_2007} describes the impact of an audience on the salience of a social category: available and visible audiences affect the expression of social identity, and the audience's feedback affects the nature of the social identity itself. Actors are particularly likely to enact group norms when they define themselves as group members and think they are visible to an audience \cite{klein_social_2007}. According to this framework, the people likely to conform to group norms most enthusiastically are those with uncertain membership \cite{branscombe1999context} who might "overconform" with an identity consolidation purpose, i.e., enacting even more intensely the group's norms to make their commitment clear \cite{klein_social_2007}.

There is evidence that the audience and its expectations regarding gender roles can shape individuals' behaviour \cite{meyers_sex_1997, postmes_behavior_2002, quayle_womens_2018}. For example, in the context of education, although both men and women can make stereotypical choices, the influence of gender stereotypes in a mixed-gender environment seems to be stronger for girls: girls tend to act more gender stereotypically in mixed-sex schools than in single-sex schools, whereas boys tend to act more gender stereotypically in single-sex schools \cite{RePEc:iza:izadps:dp7037}. These findings encourage research exploring behavioural differences between single- and mixed-gender audiences and how behaviours in these environments can reflect gender stereotypes.

\subsection{Reddit as a Platform to Study Gender Stereotypes and Audience Effects Related to Parenting}
\subsubsection{Presentation of Reddit and Parenting Subreddits}
Reddit is a social media content aggregation platform. Although headquartered in the USA, it has a global reach. Regarding demographics, Reddit users are more likely to be men, below 50 years old, and urban and suburban residents \cite{duggan_6_2013}. Content is published by pseudonymous registered members on different topical communities called "subreddits" in the form of links, text, images, or videos and can be upvoted, downvoted, or commented on by other users.  Subreddits cover topics such as politics, religion, science, video games, and many others, including parenting. The three most popular subreddits related to parenting are r/Parenting with 3,524,346 members, r/Mommit with 290,837 members, and r/Daddit with 265,191 members (as of October 2021). r/Parenting is a mixed-gender subreddit, r/Mommit is a mother-centric and overtly single-gender subreddit, and r/Daddit is a father-centric and overtly single-gender subreddit. We will focus on these subreddits to explore the extent to which parenting talk reflects common gender stereotypes, but also how it is sensitive to the audience, as we describe in more detail in the method section below.

\subsubsection{Prior Studies on Parenting Subreddits}
At least two published studies have explored the expression of parenthood on Reddit, one using an ethnographic approach \cite{feldman__2021} and the other using a computational approach \cite{ammari_pseudonymous_2018}.

Feldman's study \cite{feldman__2021} applied qualitative grounded theory to posts on r/Mommit and r/Daddit to explore how Reddit users negotiate gendered parenting ideologies. They found that Reddit users reproduce the stereotypical gender roles painting the mother as the primary caregiver and the father as the secondary caregiver on r/Daddit and r/Mommit. For instance, on r/Mommit, mothers talk about a dyadic mother-child relationship while fathers on r/Daddit talk about a triadic father-mother-child relationship. Furthermore, whereas good fathering is defined as being present and spending time with their child (shown through pictures of their child), good mothering is defined as positively shaping and impacting their child. Feldman \cite{feldman__2021} describes fathers' concerns to be centred around \textit{whether} they care for their child and mothers' to be about \textit{how} to care for their child. While Feldman's research \cite{feldman__2021} provides essential insight, the method is intensive and only allowed the analysis of one month of data. With the number of published posts on Reddit being substantial, quantitative text analysis methods such as topic modelling may allow more comprehensive analysis. Furthermore, as this work is about expressed concerns in interaction and knowing that single-gender audiences differ from mixed-gender audiences \cite{RePEc:iza:izadps:dp7037,patterson_division_2004}, it is interesting to see how Reddit parents interact on a mixed-gender subreddit. The latter can tell us whether the users differ regardless of the audience or if their expressed concerns relate to the audience they interact with. 

A study by Ammari et al. \cite{ammari_pseudonymous_2018} used topic modelling to describe the most discussed topics on r/Mommit, r/Daddit, and r/Parenting. They collected the comments published on these subreddits between 2008 and 2016, from which they identified 47 topics, summarised in eight main areas (as one area is missing in their article, we consider the seven others): (a) Birth, milestones, and transition to teen years; (b) Discipline or abuse; (c) Judgment; (d) Play and competition; (e) Purchase suggestions; (f) Faith and family; (g) Posting norms. r/Parenting users talked more about discipline and teen talk. r/Mommit users discussed the most topics such as Sleep training, breastfeeding, milestones, child weight gain, pregnancy recovery, and housework. On r/Daddit, the most frequent topics were congratulations, Neonatal Intensive Care Unit (NICU) experience, legal questions for custody battles, and Halloween costumes. Both r/Daddit and r/Mommit users talked about posting on social media, preparing food, circumcision, vaccination, and naming children.

Although Ammari et al. \cite{ammari_pseudonymous_2018} present interesting results, there are significant limitations that we will tackle in the present research. In terms of design, although they acknowledge that r/Mommit and r/Daddit are mother-centric and father-centric, respectively, they do not emphasise their particularity of being single-gender subreddits and r/Parenting being a mixed-gender subreddit, and therefore miss any possible contextual audience effects. However, research showed the effect of the audience gender at school \cite{RePEc:iza:izadps:dp7037}, online \cite{postmes_behavior_2002}, and offline \cite{meyers_sex_1997}. Therefore, the gender diversity of an audience should be considered and studied to include potential variations due to it. 

\subsection{Present Study}
The present research extends previous research to integrate the study of audience effects, a quantitative text analysis method using topic modelling, and a precise and transparent process of qualitative inference. Regarding the audience effects, little research on the differential impact of single- and mixed-gender audiences has been done, especially regarding parenting. To analyse the variation linked to the audience, participants who have interacted with both types of audiences are selected. This design allows the study of gender differences in relation to the context in which they appear. 
More specifically, this study explores how mothers and fathers discuss topics related to parenting with different audiences through the analysis of comments published in 2020 in Reddit's popular subreddits regarding parenting: r/Daddit, r/Mommit, and r/Parenting. These three subreddits mainly differ in the audience they claim to gather, either a single-gender audience (r/Daddit and r/Mommit) or a mixed-gender audience (r/Parenting). The difference in audience is assumed from the descriptions of the subreddits themselves: r/Daddit is for fathers, r/Mommit is for mothers, r/Parenting is for parents. First, we will identify mothers (users who have posted on both r/Mommit and r/Parenting but not r/Daddit) and fathers (users who have posted on both r/Daddit and r/Parenting but not r/Mommit). Then, topic modelling is used to describe parenting talk and see if the change in expressed concerns corresponds with the change of audience and if there are gender differences in relation to gender stereotypes.

\section{Method}
\subsection{Dataset}
Reddit comments were collected using Pushshift's API\footnote{Code in Python by Mikołaj Biesaga, The Robert Zajonc Institute for Social Studies, University of Warsaw. GitHub repository: https://github.com/MikoBie/reddit} \cite{baumgartner2020pushshift}, a social media data collection, analysis, and archiving platform updated in real-time. Two sets of comments were collected: (a) all the comments published in 2020 on r/Daddit and r/Parenting by authors who published on both and (b) all the comments published in 2020 on r/Mommit and r/Parenting by authors who published on both. The raw dataset contains 194,497 comments published by 8,361 unique Reddit users.

\subsection{Topic Modelling}
Topic modelling is a computer-based method developed in the field of Natural Language Processing. Given a vast amount of text data, the machine learning algorithm identifies underlying topics in the corpus. Because it is automatic, the method allows very consistent detection of topics for large datasets, whereas manual qualitative analysis, which relies on human coders, can be inconsistent \cite{carron-arthur_whats_2016} and very time-consuming. To find suitable topics, topic modelling involves counting words and grouping similar word patterns. However, topics based on word co-occurrence in documents rather than conceptual similarity can result in the algorithm's output showing hardly interpretable topics.

Latent Dirichlet Allocation \cite{blei_latent_2003} topic model is one of the most well-known topic modelling techniques. It is a generative probabilistic model of a corpus. The probabilistic model generates data by identifying how the input data was created word by word. The generated data is based on two probability distributions associated with two hyperparameters: (a) the probability \textit{q}, which captures how likely it is for each topic to represent a document. The hyperparameter ${\alpha}$, the topic prior, is associated with the probability \textit{q} and controls how many topics we expect each document to have. The higher the value ${\alpha}$, the more topics each document will be associated with. (b) the probability \textit{z}, which captures how likely it is for each word to represent each topic. The hyperparameter ${\beta}$, the word prior, is associated with the probability \textit{z} and controls how topic-specific we want each word to be. The higher the value ${\beta}$, the greater the number of topics each word will be associated with \cite{hovy_text_2020}. To test and evaluate different topic models efficiently, a coherence measure ($C\textsubscript{V}$) is used to assess the human interpretability of the model \cite{10.1145/2684822.2685324}. The interpretation and labelling of topics are qualitatively carried out using the keywords defining each topic in the output of the topic model. Once the topics are identified, we can then describe each of our documents as a distribution of those topics and find the dominant topic in each document \cite{hovy_text_2020}.

\subsection{Preprocessing}
To be coherent with our assumption that r/Mommit posters are most likely to identify as women and r/Daddit posters as men, we removed 502 authors who posted both on r/Mommit and r/Daddit as they make the comparison between these two groups less meaningful. They potentially correspond to mothers posting on r/Daddit and fathers posting on r/Mommit. Thirty-six obvious bots, identified because their pseudonyms contained the word "bot", were removed. Deleted comments, comments written by the Auto Moderator, and stop words (i.e., the words that occur in abundance and provide only little to no information about the content of textual data) were removed too. Some authors were removed because of empty comments (e.g., single-word comments such as "OK" became empty because of the removed stop words). We kept only the authors for whom the dataset still contained at least one post on a single-gender subreddit and one on the mixed-gender subreddit. In this way, we were able to analyse the effects of a change of audience for the same users. This dataset represents 145,771 comments posted by 7,791 authors (comments per author: ${\textit{M}=18.71}$, ${\textit{SD}=62.24}$, ${\textit{Mdn}=7.00}$, ${\textit{Min}=2.00}$, ${\textit{Max}=3769.00}$). Of these, users who have posted on both r/Daddit and r/Parenting in 2020 represent 2,967 users, with 16,328 comments on r/Daddit and 27,263 comments on r/Parenting; users who have posted on both r/Mommit and r/Parenting in 2020 represent 4,824 users, with 23,977 comments on r/Mommit and 78,203 comments on r/Parenting. For clarity regarding the explanation of results, we assumed that users posting on r/Daddit or r/Mommit (and not on both) identify as “fathers” and “mothers”, respectively. We made the text lower-case and removed punctuation, special characters, and hyperlinks. This dataset was kept for the second part of the study, which will use the topic model to attribute a relative frequency of topics in each comment. 

The data was lemmatised (i.e., the different inflected forms of words were replaced by their normalised form) using \textit{NLTK} \cite{BirdKleinLoper09}, a module for Natural Language Processing in Python \cite{van_rossum_python_2009} and \textit{WordNet} \cite{Fellbaum1998, 10.1145/219717.219748}, a lexical database for the English language. The data was tokenised, and the list of English stop words \cite{pedregosa_scikit-learn_2011} was removed. Additionally, after a first model inference, we added these words to the list of stop words \cite{schofield_pulling_2017} to increase interpretability: "time", "hair", "day", "month", "week", "hour", "year", "minute", "idea", "adult", "age", "comment", "people", "person", "man", "sure". The \textit{Gensim} module \cite{rehurek_lrec} was used in Python to create a bigram model for improving the quality of the inferred topics \cite{10.1145/1143844.1143967}. A bigram model uses a sliding window of two terms to examine the text (each combination of two adjacent words is considered) and find combinations of two words which should be analysed as one term because one unique concept spans the two words, e.g., "mental health" \cite{hovy_text_2020}. Using \textit{Gensim} \cite{rehurek_lrec}, a dictionary giving each term an integer identification was created, and doc2bow (i.e., document to bag-of-words) function was used to convert documents into a bag-of-words format. A document in bag-of-words format means that each document is represented by the number of unique terms it contains and the number of times each of these terms occur. To reach a more interpretable topic model, we tried to keep only nouns and nouns and verbs using \textit{NLTK}'s \cite{BirdKleinLoper09} part-of-speech tagger (i.e., a function that assigns each word a grammatical descriptor). After calculating the coherence scores for models ranging from two to 50 topics, we decided to keep the model with only nouns, resulting in higher coherence scores and greater human interpretability. The corpus used for topic modelling contained 136,256 documents (i.e., comments) and a vocabulary of 9,969 unique tokens. 

\subsection{Data Analysis}
LDA topic modelling was run\footnote{The Python code associated with this study can be found here: \url{https://github.com/melodyspr/parenting_reddit}} on the corpus using \textit{Gensim} module \cite{rehurek_lrec} in Python \cite{van_rossum_python_2009}. Coherence scores ($C\textsubscript{V}$) were used to define the optimal number of topics. $C\textsubscript{V}$ was at its highest value when the model had 12 topics. Therefore, we pursued the analysis with 12 topics. To set the hyperparameters $\alpha$ and $\beta$, we ran multiple models for different values of hyperparameters and then selected the settings that resulted in the best $C\textsubscript{V}$. The highest coherence score was ${C\textsubscript{V}=.60}$ when ${\alpha=.01}$ and ${\beta=.61}$. \textit{Chunk size} controls how many documents are processed at a time in the training algorithm, \textit{passes} control how often we train the model on the entire corpus, and \textit{iterations} control how often we repeat a particular loop over each document \cite{rehurek_lrec}. We set ${chunk size=5000}$, ${passes=40}$, and ${iterations=1000}$ and the topic model reached a coherence score  ${C\textsubscript{V}=.62}$. 

Once we have established the topic model, we can represent each comment as a distribution of topics. The keywords defining a topic were compared with the words in the comment, and from the similarity between these two sets of words, we could calculate the relative frequency of topics in each comment\footnote{Code in Python by Tawfiq Ammari, School of Information, University of Michigan. GitHub repository: \url{https://github.com/tawfiqam/Parenting-Reddit}}. Because we used a very low alpha parameter (${\alpha=0.01}$), the comments were represented only by a few topics, making the comparison between comments easier, especially for comparing groups of comments. From these relative frequencies, the average scores of topics were calculated for comments aggregated by gender and audience: mothers on r/Mommit, mothers on r/Parenting, fathers on r/Daddit, and fathers on r/Parenting.

\section{Results}
\subsection{Topics}
The resulting topic model included 12 topics: \textit{Thank you/Appreciation}, \textit{Medical care}, \textit{Education/Family advice}, \textit{Furniture/Design}, \textit{Birth/Pregnancy}, \textit{Change/Potty training}, \textit{Physical appearance/Picture}, \textit{Work/Raise children}, \textit{Food}, \textit{Leisure activities}, \textit{School/Teaching}, and \textit{Sleep training}. To label each topic, we used the list of keywords and the comments which scored the highest on the topics (see Appendix). We will give examples of keywords used to label each topic so the reader can understand the inference process:

\begin{itemize}
    \item {\texttt{Thank you/Appreciation}}: "thank", "thanks", or "advice" indicate that this topic seems to be related to parents responding to advisory or helpful posts and thanking other users.
    \item {\texttt{Medical care}}: "doctor", "issue", or "health" indicate that this topic seems to be related to parents sharing experiences about their health and their children's.
    \item {\texttt{Education/Family advice}}: "situation", "family", or "problem" indicate that this topic seems to be related to parents sharing their advice about how to handle a situation.
    \item {\texttt{Furniture/Design}}: "room", "floor", or "space" indicate that this topic seems to be related to parents discussing physical arrangements, e.g., in the house.
    \item {\texttt{Birth/Pregnancy}}: "life", "pregnancy", or "birth" indicate that this topic seems to be related to parents reacting and congratulating other parents for birth and pregnancy announcements.
    \item {\texttt{Change/Potty training}}: "diaper", "potty", or "toilet" indicate that this topic seems to be related to parents sharing experiences about their children's diaper changes, stools, and toilet training.
    \item {\texttt{Physical appearance/Picture}}: "look", "face", or "picture" indicate that this topic seems to be related to parents reacting to pictures and physically describing their children.
    \item {\texttt{Work/Raise children}}: "home", "family", or "work" indicate that this topic seems to be related to parents sharing experiences about finding a balance between work and raising children.
    \item {\texttt{Food}}: "food", "meal", or "snack" indicate that this topic seems to be related to parents sharing recipes and experiences about their children eating.
    \item {\texttt{Leisure activities}}: "book", "play", or "fun" indicate that this topic seems to be related to parents sharing about their free-time activities.
    \item {\texttt{School/Teaching}}: "school", "teacher", or "class" indicate that this topic seems to be related to parents sharing about academic education and teaching practices.
    \item {\texttt{Sleep training}}: "night", "sleep", or "bedtime" indicate that this topic seems to be related to parents sharing about sleep training.
\end{itemize}

\subsection{Relative frequency of topics}
Figure 1 shows topics' scores (i.e., the average relative frequency of topics in comments) for the different groups (i.e., fathers on r/Daddit, fathers on r/Parenting, mothers on r/Mommit, and mothers on r/Parenting).

\begin{figure}[htbp]
  \centering
  \includegraphics[width=\linewidth]{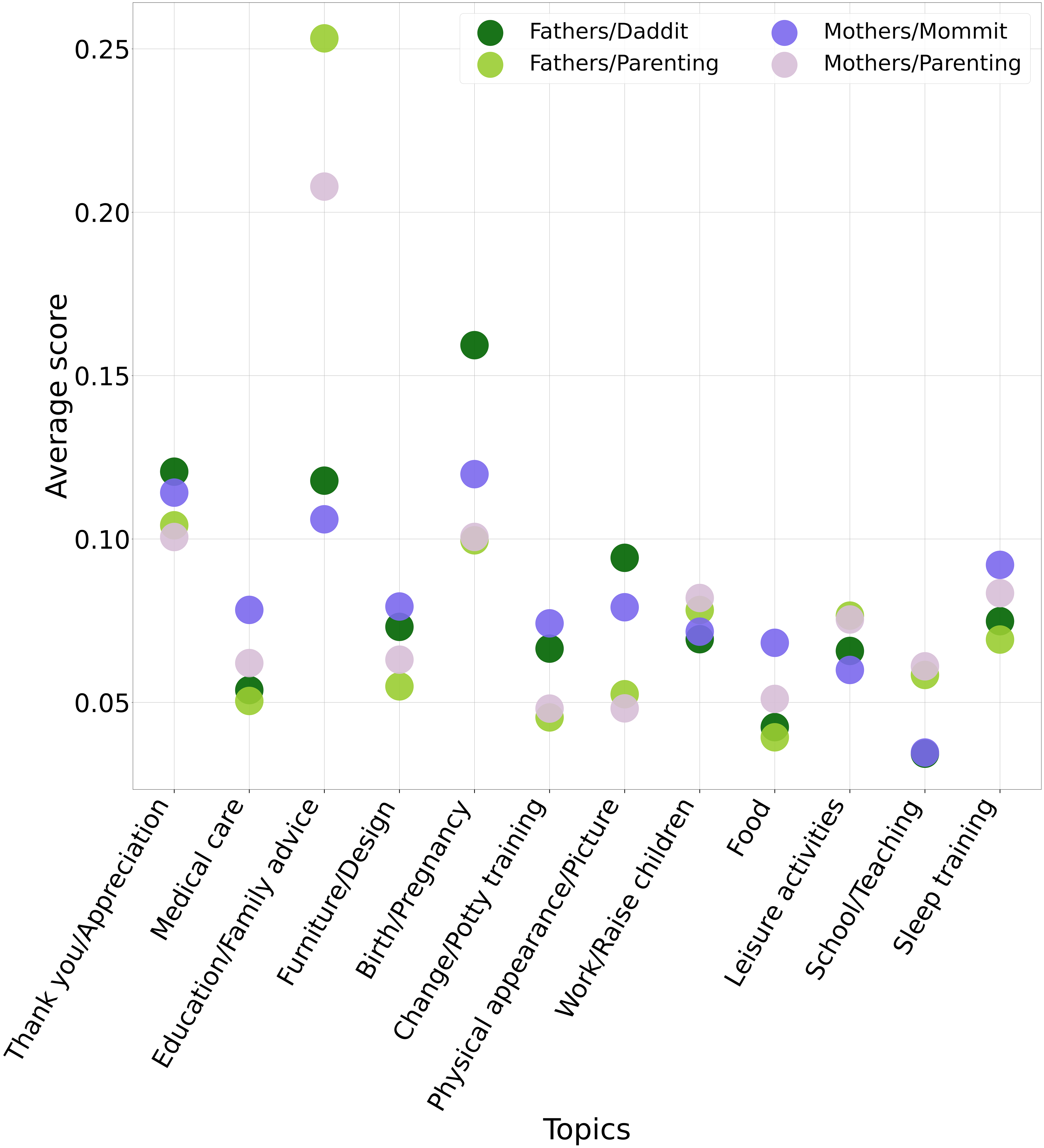}
  \caption{Scores of mothers and fathers' comments on the twelve topics}
  \Description{The figure is showing the scores of each group (fathers on r/Daddit, fathers on r/Parenting, mothers on r/Mommit, and mothers on r/Parenting) for each topic. As the figure is best described with a table, we added a table in the Appendix (Table 2) which shows the same data.}
\end{figure}

We will describe the results and analyse the topics summarised into four areas of concern: Basic needs (\textit{Food}, \textit{Sleep training}, \textit{Medical care}, \textit{Change/Potty training}), Announcements and introductions (\textit{Birth/Pregnancy}, \textit{Thank you/Appreciation}, \textit{Physical appearance/Picture}), Education and leisure (\textit{Education/Family advice}, \textit{Work/Raise children}, \textit{School/Teaching}, \textit{Leisure activities}), and Physical arrangements (\textit{Furniture/Design}). The following four paragraphs will develop each of these areas.

Basic needs were more discussed by mothers. Except for \textit{Change/Potty training}, mothers discussed more than fathers the topics related to children's basic needs, regardless of the audience. Mothers (on r/Mommit and r/Parenting) discussed more the topics \textit{Sleep training}, \textit{Food}, and \textit{Medical care} than fathers (on r/Daddit and r/Parenting). Regarding \textit{Change/Potty training}, there was a clear distinction between the mixed-gender subreddit and the single-gender ones; the topic was more discussed on the latter. However, although the difference seemed to be due to the audience, mothers on r/Mommit discussed \textit{Change/Potty training} more than fathers on r/Daddit and mothers on r/Parenting more than fathers on r/Parenting. Among mothers, a similar interaction between gender and audience was visible in the other topics related to basic needs as mothers always discussed them more on the single-gender subreddit r/Mommit.

Announcements and introductions were an important topic for fathers in the single-gender subreddit. Fathers on r/Daddit had the highest scores on the topics \textit{Birth/Pregnancy}, \textit{Thank you/Appreciation}, and \textit{Physical appearance/Picture}. Among these topics, the topic \textit{Birth/Pregnancy}, which includes birth announcements and congratulating comments, showed the greatest difference between fathers' comments on the single-gender subreddit r/Daddit and other comments.

Education and leisure were especially discussed by mothers and fathers on r/Parenting. In other words, the topics \textit{Education/Family advice}, \textit{Work/Raise children}, \textit{Leisure activities}, and \textit{School/Teaching} were more discussed with the mixed-gender audience. These topics seemed to represent norms regarding the expected topics Reddit users should discuss on r/Parenting. Regarding gender differences, mothers and fathers discussed \textit{Work/Raise children}, \textit{Leisure activities}, and \textit{School/Teaching} similarly, whereas \textit{Education/Family} advice was discussed more by fathers than mothers. \textit{Education/Family advice} was also the most discussed topic overall. 

Physical arrangements were more discussed in the single-gender subreddits than in the mixed-gender. 

Findings showed gender differences and differences between single and mixed-gender subreddits, but, for some results, the subreddits' norms and their potential influence should be considered. Our sample included four groups (i.e., fathers on r/Daddit, fathers on r/Parenting, mothers on r/Mommit, mothers on r/Parenting) but only three subreddits. This fact was, to some extent, reflected in the results. Across all topics, there seemed to be more similarities between mothers and fathers on r/Parenting than between other groups. For instance, there appears to be an important overlap between mothers and fathers on r/Parenting for the topics \textit{Thank you/Appreciation}, \textit{Birth/Pregnancy}, \textit{Change/Potty training}, \textit{Physical appearance/Picture}, \textit{Leisure activities}, and \textit{School/Teaching}.

\section{Discussion}
The findings of the study illustrate the concerns expressed by mothers and fathers on Reddit. Leaning on the social identity performance framework that shows the importance of an audience for social identity salience \cite{klein_social_2007}, the study aimed to explore gender differences in relation to an audience. Three popular subreddits related to parenting were selected as they differ in the gender and audience they target and attract: r/Daddit is a father-centric subreddit with an overtly single-gender audience, r/Mommit is a mother-centric subreddit with an overtly single-gender audience, and r/Parenting is a parent-centric subreddit with a mixed-gender audience. Parents' posts across these subreddits were analysed using LDA topic modelling, which captured 12 topics summarised into four areas: Basic needs (\textit{Food}, \textit{Sleep training}, \textit{Medical care}, \textit{Change/Potty training}), Announcements and introductions (\textit{Birth/Pregnancy}, \textit{Thank you/Appreciation}, \textit{Physical appearance/Picture}), Education and leisure (\textit{Education/Family advice}, \textit{Work/Raise children}, \textit{School/Teaching}, \textit{Leisure activities}), and Physical arrangements (\textit{Furniture/Design}). These 12 topics were then used to compare parents in terms of gender and audience, resulting in three main findings. First, basic needs were more discussed by mothers regardless of the audience. Second, announcements and introductions were more discussed by fathers with a single-gender audience. Third, education and leisure were discussed by both mothers and fathers, especially with a mixed-gender audience. We will put these results into perspective, comparing them with theories and previous studies in the following paragraphs. 

Firstly, findings showed gender differences between mothers and fathers. The latter appeared in three topics, with mothers always discussing them the most: \textit{Food}, \textit{Sleep training}, and \textit{Medical care}. Sleeping, eating and health are probably the most important topics related to the child's survival; they represent their basic needs. By being the users who discuss the most these topics, mothers position themselves (and are positioned) as the primary caregiver with communal characteristics \cite{eagly_gender_1984, ellemers2018gender}. These findings are coherent with the previous studies of Ammari et al. \cite{ammari_pseudonymous_2018} and Feldman \cite{feldman__2021}, conducted in the same context. Ammari et al.'s \cite{ammari_pseudonymous_2018} found that r/Mommit users are the ones discussing the most topics related to sleep training, breastfeeding, milestones, child weight gain, and pregnancy recovery. Feldman's analysis \cite{feldman__2021} showed mothers as more concerned about how to care for their child, i.e., child-rearing's practical and concrete sides. Therefore, gender differences, as predicted by previous research, were reflected in the present study, especially for mothers expressing concerns about the child's basic needs, and this was the case regardless of audience and context for the topics \textit{Food}, \textit{Sleep training}, and \textit{Medical care}.

Secondly, findings partially demonstrated an audience effect as described by previous research. Research on the Self-Categorization Theory \cite{turner_rediscovering_1987} and the social identity performance \cite{klein_social_2007} has shown how the salience of a social category and the audience result in individuals behaving in coherence with the group norms. Additionally, as demonstrated in the comparison between mixed and single-sex schools, the mixed-gender environment pushed girls towards more stereotypical behaviours than the single-gender environment, whereas the opposite happened for boys \cite{RePEc:iza:izadps:dp7037}. In the present study, it seems like fathers did act more gender stereotypically with a single-gender audience, but for mothers, the findings are more nuanced and depend on the topic. With single-gender audiences, both mothers and fathers announced pregnancies and births and introduced their children. However, fathers, in terms of relative frequency, scored higher on these topics than mothers. The latter confirms Ammari et al.'s \cite{ammari_pseudonymous_2018} findings where one of the most discussed topics in r/Daddit was congratulations. The results are also coherent with Feldman's \cite{feldman__2021} results: r/Daddit users talking about \textit{whether} they care for their child through posting pictures, announcing pregnancies and births, and reacting to these announcements. In the present work, the topics \textit{Thank you/Appreciation}, \textit{Birth/Pregnancy}, and \textit{Physical appearance/Picture} touch on an abstract, superficial aspect of parenting, which is more about the recognition of a "parent" identity (the \textit{whether} \cite{feldman__2021}) than the actual parenting function (the \textit{how} \cite{feldman__2021}). This quest for recognition and the emphasis on physical appearance and celebration create an almost trophy-like image of the child and can be associated with gender stereotypes related to the expectation of performance for men \cite{ellemers2018gender} and confirm the increase of gender-stereotypical behaviours for fathers in single-gender environments \cite{RePEc:iza:izadps:dp7037}. The results are more nuanced for mothers. On the one hand, mothers expressed less stereotypical concerns with a single-gender audience, as they talked more about topics related to recognition and celebration. On the other hand, mothers also expressed more stereotypical concerns with a single-gender audience. Indeed, mothers expressed concerns regarding the child's basic needs (\textit{Sleep training}, \textit{Food}, and \textit{Medical care}) preferentially with a single-gender audience. Therefore, although the audience seemed to have an expected effect on fathers with announcements and introductions, the findings were unexpected for mothers who discussed more the stereotypical area of basic needs with a single-gender audience than with a mixed-gender audience.

Thirdly, under pseudonyms, fathers on r/Parenting interact with other fathers and with mothers, from whom they may wish to gain recognition as legitimate parents. The salience of the "parent" social category and the threat to be rejected because they are expected to be secondary parents \cite{wall_how_2007} can push them towards "overconformism" \cite{klein_social_2007}, making them behave more as prototypical parents than mothers. The topic \textit{Education/Family advice} seems to reflect this phenomenon. This topic is discussed approximately twice as much with the mixed-gender audience than with the single-gender audiences by both mothers and fathers. However, although both mothers and fathers posted a lot about this topic, the difference between mothers and fathers on r/Parenting is the greatest among all topics. The latter might reflect fathers' overconformity in this context. As an alternative explanation, we can also question the particularity of the topic \textit{Education/Family advice}. The comments which scored the highest on this topic reflect an agentic aspect of parenting \cite{eagly_gender_1984} where parents give advice about what they would do and what other parents should do in different situations. This agency seems to reflect the position of authority or control fathers are associated with \cite{10.2307/3115186} and be, thus, a stereotypical behaviour associated with men. Therefore, fathers talking predominantly about education and giving family advice with the mixed-gender audience is difficult to interpret as it can be due to overconformism but also to the agency associated with advising others.

\subsection{Implications}
The present work explored how expressed concerns about parenting can inform us of gender stereotypes and audience. Additionally, this study introduced a novel method to explore, in parallel, differences between two groups (i.e., mothers and fathers) and individuals' contextual differences (i.e., with a single-gender audience or a mixed-gender audience). This method allowed the capture of gender and audience differences regarding parenting talk on Reddit. With this method, mothers and fathers' behaviours were analysed when separated in single-gender audiences and gathered in a mixed-gender audience. Previously, in a study gathering r/Daddit, r/Mommit, and r/Parenting, Ammari et al.'s \cite{ammari_pseudonymous_2018} had included r/Parenting comments in their dataset without acknowledging the potential impact induced by the change in audience. Leaning on the social identity performance framework \cite{klein_social_2007} and Favara's study \cite{RePEc:iza:izadps:dp7037} about single-sex schools, the inclusion of r/Parenting was made meaningful and allowed the exploration of audience and gender in interaction. In the following two paragraphs, we will expand on the scientific contribution of the present research and its potential societal impact. 
The findings contributed to two research areas: the expression of concerns related to gender stereotypes and the relationship between audience change and behavioural change. 

The results presented here enrich the literature about the evolution of gender roles by providing a picture of parents' concerns expressed on Reddit in 2020. The topics and their distribution across users' comments showed a gendered division of parenting in parents' discourse. Additionally, the research supported the hypothesis of a gender-differentiated impact of mixed and non-mixed audiences on individuals. Research on this differential impact has been mainly studied in the education context \cite{RePEc:iza:izadps:dp7037} despite potentially touching every sphere of society. Therefore, we broadened the research on single-gender and mixed-gender environments by applying this perspective in the social media context in relation to parenting. To suggest a more robust methodology, we included the framework of social identity performance \cite{klein_social_2007}. This way, we focused on individuals' behavioural changes related to an audience as a social identity activator. The integration of the two frameworks of understanding, i.e., gender stereotypes and audience, allowed for a holistic approach to the analysis of expressions online. 

Regarding the societal impact of this research, the findings echoed the unequal division of labour in the offline world. On the one hand, mothers expressed concerns about the practical aspects of child-rearing, reflecting the characteristics associated with motherhood. On the other hand, fathers seem to be more concerned about expressing their fatherhood and showing that they care for their child. Although this study does not establish any causal relationship between stereotypes and gender differences on Reddit and only describes their overlap, we can still question the potential effect of stereotypes. Stereotypes can profoundly impact behaviours, resulting in group-based expectations that restrain individuals from behaving in a non-expected manner. In this study, the overlap between parenting talk and gender stereotypes shows how stereotypes constitute a powerful framework for looking at and understanding interactions regarding parenting.

\subsection{Limitations}
\subsubsection{A non-representative sample}
The sample does not include all parents who published on the three subreddits in 2020. We have only selected mothers and fathers who have published comments on both a single-gender subreddit (r/Mommit or r/Daddit) and a mixed-gender subreddit (r/Parenting). This participants' selection strategy might have biased the results. Indeed, parents may choose to post comments on two different subreddits about parenting precisely because they want to express different concerns to different audiences. Therefore, our results might not apply to parents who published on only one of the subreddit. For example, r/Daddit users who publish exclusively on this subreddit might express all their concerns (including all the topics found in this study) to that audience. Thus, the choice of reducing the sample to only the users who published on at least two parenting subreddits may have biased the findings.

The sample also has an impact on the generalisability of the study. Reddit data is not representative: Reddit parenting communities do not represent all social media parenting communities, let alone offline parenting. For instance, Reddit is popular in English-speaking countries. Even if there are other languages on Reddit, the subreddits we chose gathered almost exclusively comments written in English. Therefore, the work cannot be generalised to all cultures. Moreover, fathers writing on r/Daddit and sharing experiences about fatherhood may be a population that is already more concerned about parenting than the average father. Therefore, the latter can result in fewer differences between mothers and fathers on Reddit (in these subreddits) than there is offline. Thus, the particularity of Reddit and its users should be considered to prevent overgeneralisation.

\subsubsection{Analytic judgement}
We note that the advanced techniques of machine learning and topic modelling do not erase interpretation bias. Once the topic model has been created, the researcher has several sets of words from which they want to make sense. Although some topics are almost obvious, others are more subject to interpretation, and this judgement is qualitative.

\subsection{Future Research}
The present research has shown how parenting talk reflects gender stereotypes and how the audience can shape the expression of concerns. To support and solidify the findings about the audience effect, the gendered expression of concerns and an actual unequal division of labour between parents, this exploratory and observational study can be complemented by an online experiment and a survey. With an online experiment, we can effectively anonymise participants and make them interact online with other participants about parenting, either with a single-gender audience or with a mixed-gender audience. The gender identifiability, the name of the communities, and the number of mothers and fathers can be manipulated. With a survey, it will be possible to explore the relationship between online and offline behaviours related to parenting. Finally, for further investigation, we would like to question the use of gender as binary in this study and encourage future research to consider other possible definitions of gender and parenting, which can be different from the simple spectrum of motherhood and fatherhood.

\section{Conclusion}
In the present investigation, we studied parenting talk on Reddit to explore how it reflects gender stereotypes and is sensitive to the audience. Findings indicated that, first, mothers posted more about children's basic needs, regardless of the audience, reflecting the communal characteristics stereotypically associated with women. Secondly, fathers posted more than mothers to announce pregnancies and births and introduce their child, especially to a single-gender audience, showing a more superficial aspect of parenting focused on the social recognition of the "father" identity. Finally, among Reddit parents, discussing education and giving advice were mostly expressed to a mixed-gender audience and more by fathers than mothers, reflecting either the agentic characteristics associated with masculinity or overconformism. The study of audience and gender stereotyping in parenting talk gives an important window into the construction of gender identity in private contexts, which are usually out of view.

Although no link can be directly drawn between these findings and offline behaviours, the echo of gender stereotypes in these online interactions questions the impact of stereotypes in the real world. When mothers decide to behave in a stereotype-based expected manner, they usually have to fragilise their careers through less demanding or part-time jobs. These stereotypes are consequential. Child-rearing not being a paid occupation, instantiating a stereotypically appropriate parenting identity imposes a professional opportunity-cost for women: if women are expected to be primary caregivers at home with their children, men are expected to be outside and contributing to society's organisation.

\begin{acks}
We would like to thank Geoffrey Cideron for his useful suggestions throughout this work. This publication has emanated from research conducted with the financial support of Science Foundation Ireland under Grant number 18/CRT/6049 and the European Research Council (ERC) under the European Union's Horizon 2020 research and innovation programme (grant agreement No. 802421).
\end{acks}

\bibliographystyle{ACM-Reference-Format}
\bibliography{main}

\appendix
\newcommand{\specialcell}[2][c]{%
 \begin{tabular}[#1]{@{}c@{}}#2\end{tabular}}
 
\section{Topics’ Labels, Keywords, and High-Scoring Comments}
%% The text in cells should be left-aligned and centered. + the table should be split on two pages
\begin{table*}[htp]
%\begin{sidewaystable}
    \centering
  \caption{Topics’ labels, keywords, and high-scoring comments}
  \label{tab:topics}
  \begin{tabular}{p{0.15\textwidth}p{0.2\textwidth}p{0.6\textwidth}}
    \toprule
    Label &Keywords & High-scoring comments\\
    \midrule
    %\multirow{1}{*}[\normalbaselineskip]
    {\texttt{Thank you/appreciation}} & %\specialcell{
    "thank", "word", "thanks", "toddler", "sound", "advice", "today", "moment", "hug", "mama"%}
    & %\specialcell{"
    "Thank you! I appreciate your insight and you made good points especially on reversing the words I'm using too:) I like it. Thank you and I really appreciate it."%}
    \\
    %\multirow{1}{*}[\normalbaselineskip]
    {\texttt{Medical care}} & %\specialcell{
    "doctor", "issue", "hospital", "formula",  "anxiety", "pain", "risk", "depression",  "health", "weight"%}
    & %\specialcell{
    "I have the Copper. I can't do the hormonal ones,not even the minipill. Severe Migraines and doctor feared a higher chance of clots. I've had spotting between periods, but only 2 periods in 14mos. They are very painful.I should note, I didn't find childbirth painful and didn't need pain meds afterwards so I'm not that sensitive to pain."%}
    \\
    \texttt{Education/Family advice} & %\specialcell{
    "situation", "life", "family", "relationship", "problem", "issue", "need", "behavior", "choice", "question"%}
    &
    %\specialcell{
    "If I were you, I would call CPS on your husband. I called CPS on my ex while we were still together.He needs tough boundaries and consequences for his choices. You need to protect your children. CPS can provide you with resources to help, and can provide him with resources if he chooses to seek help. Honestly, unless he stops drinking your relationship is over anyways. Set yourself up to be the custodial parent by doing the right things to protect your kids well-being."%}
    \\
    \texttt{Furniture/Design} & %\specialcell{
    "room", "toy", "floor", "door", "clothes", "house", "head", "bathroom", "foot", "space"
    %}
    & %\specialcell{
    "Or, stand on a rug,l and draw an outline of yourfeet. Cut them out and duck tape them to your feet.Plus, you never need to buy carpets for your house"%}
    \\
    \texttt{Birth/Pregnancy} & %\specialcell{
    "life", "friend", "love", "partner", "pregnancy", "family", "heart", "birth", "care", "sorry"%}
    & %\specialcell{
    "Congrats dad! I'm pretty convinced that when babies come out this big they're a lot happier and easier than the littler ones, so I hope that's true for you!"%}
    \\
    \texttt{Change/Potty training} & %\specialcell{
    "diaper", "post", "body", "hand", "sex",  "potty", "poop", "toilet", "arm", "grandma"%}
    & %\specialcell{
    "I've been changing her constantly. I typically cloth diaper but I have her in honest co diapers and I'm changing these things 2-3 times an hour!"%} 
    \\
    \texttt{Physical appearance/Picture} & %\specialcell{
    "dog", "look", "face", "eye", "picture", "cat", "memory", "stranger", "color", "cousin"%}
    & %\specialcell{
    "My son is 6 months old and he has long eye lashes like in Incredibles coz he looks just like that"%}
    \\
    \texttt{Work/Raise children} & %\specialcell{
    "home", "family", "work", "house", "car", "job", "money", "friend", "seat", "birthday"%}
    & %\specialcell{
    "Raising a kid takes a village. Give yourself a break on the weekends or after work if you can. Go for a walk before you pick him up, and get some help so you can have some time. Everyone needs personal time."%}
    \\
    \texttt{Food} & %\specialcell{
    "food", "water", "meal", "milk",  "snack", "dinner", "bottle", "cup", "lunch", "eat"%}
    & %\specialcell{
    "One day I was in the kitchen cooking dinner and I turned my back for 1 second. I turn back and my son had nearly eaten a full clove of garlic like it was a piece of candy. Kids are crazy lol"%}
    \\
    \texttt{Leisure activities} & %\specialcell{
    "book", "game", "play", "attention",  "activity", "phone", "tv", "video", "fun", "stuff"%}
    &
    %\specialcell{
    "Indeed, board games gets them really excited :) Pick the games that is more about the adventure / fun than on the winning / losing, so that the focus is not so much on that "end" part.Thanks for the great suggestions :)"%}
    \\
    \texttt{School/Teaching} & %\specialcell{
    "school", "teacher", "class", "daycare",  "home", "work", "language", "group", "college", "therapy"%}
    & %\specialcell{
    "I've heard that teaching multiple languages might delay your child's spoken language a little bit compared to just teaching them one language but it's only temporary and the pay off is they end up fluent in more than one language."%}
    \\
    \texttt{Sleep training} & %\specialcell{
    "night", "bed", "sleep", "morning", "room", "nap", "bedtime", "crib", "work", "schedule"%}
    & %\specialcell{
    "I am definitely a morning person! And I struggle to get back to sleep, too. I just like very peaceful, easy mornings. My daughter (who has always been my champ sleeper) takes after me. My son was the baby that woke up every 2h on. the. dot. until I finally had it in me to sleep train after he turned 1. He wakes up ready to charge and tends to be my morning mayhem instigator, despite his efforts to not."%}
    \\
    \bottomrule
  \end{tabular}
%\end{sidewaystable}[htdp]
\end{table*}

\section{Scores of the Groups for Each Topic (Complement to Figure 1)}
\begin{table*}[htp]
  \caption{Scores of the groups for each topic (complement to Figure 1)}
  \label{Figure description figure 1}
  \begin{tabular}{lcccc}
    \toprule
    Topics &Fathers/Parenting& Fathers/Daddit & Mothers/Mommit & Mothers/Parenting\\
    \midrule
    \texttt{Thank you/appreciation} & .104&.121&.114&.101\\
    \texttt{Medical care} & .050&.054&.078&.062\\ 
    \texttt{Education/Family advice} & .253&.118&.106&.208\\    
    \texttt{Furniture/Design} & .055&.073&.079&.063\\    
    \texttt{Birth/Pregnancy} & .100&.159&.120&.101\\    
    \texttt{Change/Potty Training} & .045&.066&.074&.048\\    
    \texttt{Physical appearance/Picture} & .053&.094&.079&.048\\    
    \texttt{Work/Raise children} & .078&.069&.072&.082\\    
    \texttt{Food} & .039&.042&.068&.051\\    
    \texttt{Leisure activities} & .077&.066&.060&.075\\    
    \texttt{School/Teaching} & .058&.034&.035&.061\\    
    \texttt{Sleep training} & .069&.075&.092&.083\\    
    \bottomrule
  \end{tabular}
\end{table*}

\end{document}